# Room-temperature continuous-wave upconversion white microlaser in a rare-earth doped microcavity

Bo Jiang[1], Song Zhu[1,2], Wenyu Wang[1], Lei Shi[1, *] and Xinliang Zhang[1]

[1]Wuhan National Laboratory for Optoelectronics, Huazhong University of Science and Technology, Wuhan 430074, China

[2]School of Electrical and Electronic Engineering, Nanyang Technological University, 50 Nanyang Avenue, 639798 Singapore

correspondence to: lshi@hust.edu.cn

**KEYWORDS:** White microlaser, Whispering gallery mode, Microcavity, Rare earth, Upconversion

**ABSTRACT:**

White laser covering the visible color spectrum is critical in various applications including vivid display, holographic imaging and light-based version of Wi-Fi, but it is still challenging to realize the white microlaser due to the rigorous requirement in the balance between the optical gain and the feedback at different wavelengths. Here, we employ Tm, Er and Yb ions corporately for the upconversion white lasing in a single ultrahigh quality (Q) (up to $10^8$) doped microcavity, where the thresholds of the red, green and blue lasers are about 90, 500 and 300 μW, respectively. To the best of our knowledge, it is the first rare-earth elements based room-temperature, continuous-wave white microlaser, which exhibits relatively stable chromaticity over 180 minutes, making it possible for practical applications.

**INTRODUCTION:**

White lasers have the potential used for fluorescence microscopy[1-3], color lighting and display[4-6], as they have the higher power density, the higher contrast ratio and the wider color gamut[5], compared with the traditional light emitting diodes. Besides, as multi-wavelength coherent light sources, white lasers can also be applied in optical signal processing[7-9], high-speed visible-light communications[8-10] and holograms[9, 11]. However, the realization of white lasers is still suffering from the balance between the optical gain and the feedback at different wavelengths and the compatibility between gain materials and cavity structures[12]. Some works utilized discrete devices to produce three primary colors, which increase the volume, the complexity and the cost of the overall system[5, 13]. Further white lasers in a single device have been realized by quantum dots[12], dye doped polymers[5, 10, 14-16] and gas[17]. Remarkably, Ning et al. integrated the ZnCdSe alloy on a monolithic chip, achieving white laser under 355-nm pulsed laser pump[18].

Rare-earth (RE) elements with an abundant long-lived energy-level configuration can achieve visible emission by upconversion and downshifting processes[19]. Besides, RE elements show advantages of low-cost fabrication, high environmental stability and high photoluminescence quantum yield[20]. Therefore, RE elements are widely applied in a variety of areas including three-dimensional display[21, 22], luminescent biomarkers[19, 23, 24], super-resolution nanoscopy[25, 26] and laser fabrication[20, 27, 28]. RE based upconversion white laser was firstly realized in $NaYF_4$ hexagonal microrods under 980-nm nanosecond-pulse laser excitation[1]. Subsequently, upconversion white random lasing in $NaYF4:Yb/Er/Tm@NaYF4:Eu$ core-shell nanoparticles under 980-nm continuous-wave (CW) laser pump was realized[29]. However, the optical feedback for random lasers is based on optical scattering, which usually makes it more unstable and have the higher lasing thresholds, compared with the laser based on the optical reflection feedback.

Ultra-high-quality (Q) whispering-gallery-mode (WGM) cavities are excellent platforms for achieving stable, low-threshold and narrow-linewidth lasers[30-35]. In this demonstration, we fabricated an Er-Tm-Yb co-doped microsphere cavity with an ultrahigh intrinsic quality factor of $1.14×10^8$, achieving white upconversion laser under room-temperature and CW pump, in which the thresholds of red, green and blue (RGB) lasers are about 90, 500 and 300 µW, respectively. To the best of our knowledge, it is the first rare-earth CW white microlaser under room temperature. Besides the ultra-low threshold, this white microlaser also exhibits stable chromaticity more than 180 minutes, indicating its potential for practical applications.

**Result and discussion**

All the experiments were performed under room temperature and CW pump. As shown in Figure 1a, a Tm-Er-Yb elements codoped microsphere cavity with a diameter of about 53 µm is coupled with a microfiber with a diameter of about 1.2 µm. Three primary RGB lasers with an appropriate ratio of their intensities are achieved simultaneously pumped by a 975 nm diode laser, indicating a desired white laser. The detailed emission processes of three primary colors are shown in Figure 1b. Yb ions are used as the sensitizer to improve the upconversion efficiency in Er and Tm ions[36, 37]. The presence of blue emission is mainly contributed by the transition of $^1D_2 \to \ ^3F_4$

(455 nm, Tm), $^1G_4 \to {}^3H_6$ (480 nm, Tm) and $^4F_{3/2}, {}^4F_{5/2} \to {}^4I_{15/2}$ (450 nm, Er). The presence of green

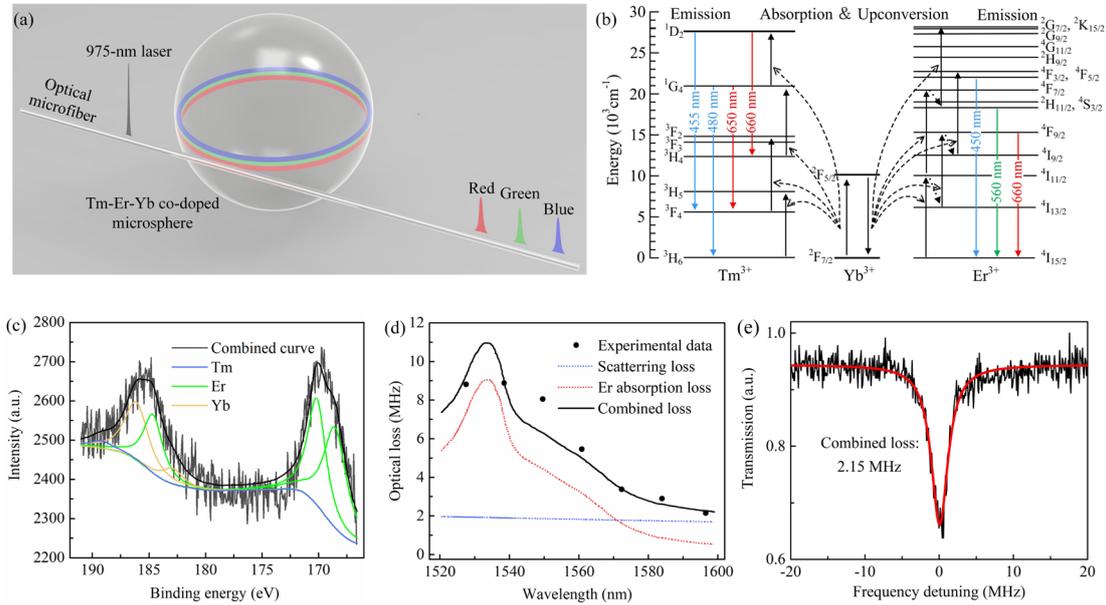

**Figure 1.** a) Schematic diagram of CW upconversion white lasing in an Tm-Er-Yb co-doped microsphere cavity under 975-nm CW laser excitation. b) Energy level diagram and proposed RGB light emission mechanisms. The dashed and dotted arrows represent ET and multi-phonon relaxation processes, respectively. The upward full arrows represent the upward transition induced by photon excitation and ET processes. The downward full arrows represent the downward transition induced by photon emission and ET processes. c) XPS spectrum characterization of the doped microcavity. d) Optical loss evolution in a wavelength range covering six FSRs, the black fitting curve represents gain materials absorption loss combined with surface scattering loss. e) Q factor characterization of the device at 1596 nm. The black and red lines represent the experimental measurement and the theoretical fitting, respectively.

emission is mainly contributed by the transition of $^2H_{11/2}, {}^4S_{3/2} \to {}^4I_{15/2}$ (550 nm, Er). The presence of red emission is mainly contributed by the transition of $^1G_4 \to {}^3H_6$ (650 nm, Tm), $^1D_2 \rightarrow {}^3H_4$ (660 nm, Tm) and $^4F_{9/2} \rightarrow {}^4I_{15/2}$ (660 nm, Er). In this scheme, green and red emission refer to the two-photon upconversion process, and blue emission refers to the three-photon upconversion process. The upconversion process is mainly comprised by excited state absorption (ESA) and energy transfer (ET) from Yb sensitizers to Tm and Er emitters. As the ET process is enhanced by the doping concentration, a higher doping concentration is needed to achieve a higher upconversion efficiency. By utilizing Guassian-Lorentzian fitting, the Er 2o3, Er 4d3, Yb 4d5 and Yb 2o3 peaks are clearly shown in Figure 1c. The doping concentration of Tm is only a fiftieth of that of Er ions, thus the peaks from Tm are inconspicuous. For this doped microsphere, the total optical loss is mainly contributed by the gain materials induced absorption loss and the cavity induced surface scattering loss. Surface scattering loss of the microcavity is approximately equal to intrinsic loss of microcavity, as the absorption loss induced by silica and radiation loss should be ignored. Here we measured the total optical loss of a certain resonance mode over six FSRs (free spectrum ranges) with a spacing of one FSR (about 11 nm) by a tunable laser. In order to avoid the thermal nonlinear effect in the microsphere cavity and the Er saturated absorption, the sweeping speed is set to 4.8 GHz/ms and the pump power is below 10 μW. The experimental results are well matched with the absorption loss of Er at 1550 nm band[38] combined with the surface scattering loss of the cavity[39], as shown in Figure 1d. It can be found that the Er absorption loss is about 9 MHz at 1535 nm, therefore, the effective doping concentration of Er is estimated to be $0.955 \times 10^{17}/cm^3$. And also, according to the doping ratio, the effective doping concentration of Tm and Yb ions are estimated to be $0.191×10^{16}$ and $0.764×10^{17}$/cm$^3$, respectively (The details are shown in Supplementary Materials). The fitted surface scattering loss at 1596 nm is about 1.7 MHz, which indicates an ultrahigh intrinsic Q factor of $1.14×10^8$. The transmission spectrum around 1596 nm is also shown in Figure 1e. As the shorter the wavelength is, the stronger the rayleigh scattering loss is, homogeneous RE-ion-doping for maintaining the ultra-high Q factor plays an essential role in achieving low-threshold and narrow-linewidth upconversion lasers.

In this nonresonant pump scheme, the pump laser with a linewidth of ~1 nm covers many resonant wavelengths of the microsphere cavity, thus a lot of WGMs will be excited to achieve the multi-mode laser. As the competition among lasing modes is unpredictable, and the lasing modes cannot be resolved clearly, the output intensity is characterized as the integration of the lasing band instead of a single lasing mode. Figure 2 shows the log-log plots of the output intensity versus the pump power. The superlinear curve is well-known S-like shape. As the pump power increases, the curve shows the transition from spontaneous emission to stimulated emission, and eventually tends to become saturated. The abrupt increase of the slope corresponds to the lasing threshold. By fitting the curves, the thresholds of red, green and blue lasers are estimated to be 90, 500 and 300 µW, respectively. As the resonant pump is more efficient than the nonresonant pump[34] (32 µW[40] for Tm-ion-based blue lasing, and 690 µW[35] for Er-ion-based green lasing), these thresholds are still very low for the upconversion microlaser under nonresonant pump. Among of them, the green laser has the highest threshold, and the red laser has the lowest threshold. In this scheme, the gain material induced absorption loss is dominated around 1550 nm, and the microcavity is weakly coupled with the microfiber at upconversion bands. Therefore, the gain material induced absorption loss should be also dominated at upconversion emission bands, as the lifetimes of the upper states are much shorter than that of the $^4I_{13/2}$ state, while the absorption cross-section is inversely proportional to the lifetime of the energy level. Once the populations are inversed to overcome the gain material induced absorption loss, the net optical gain of a small signal should be close to be positive for achieving lasing (For the detailed analysis, see the Supplementary Materials). During the green lasing process, its intermediate-energy-level lifetime is much shorter than that of red laser, which means that the $^2H_{11/2}$ and $^4S_{3/2}$ states demand higher pump powers to achieve population inversion, leading to a higher threshold. Besides, blue laser should take advantage of the cross-relaxation (CR) based energy-looping phenomenon for a lower threshold[41]. The differences among the thresholds of RGB lasers will cause an obvious shift of the chromaticity under a low pump power, at which the chromaticity will shift from red toward blue, and then toward green. The insets show that the variation of the relative emission intensity by rotating the polarizer from 0 to 180 degrees with a step size of 10 degrees. It can be found that the relative emission intensities of RGB lasers are symmetric with respect to 90 degree. Among them, the green laser is completely linearly polarized, while the others consist of transverse electric (TE) and transverse magnetic (TM) modes.

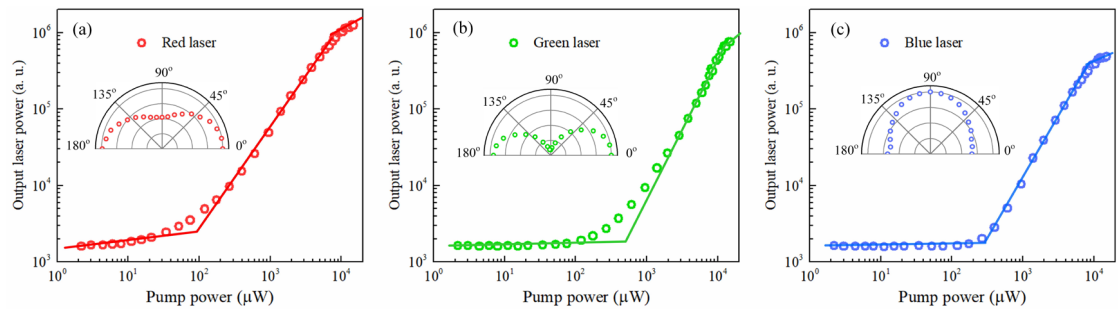

**Figure 2.** a, b, c) Emission intensities of red (a), green (b) and blue (c) lasers versus the pump power in log-log scale, respectively. Their emission intensities are integrated from 619 to 700 nm, 514 to 577 nm, and 446 to 495 nm, respectively. Inset: the emission intensity as a function of the polarization angel.

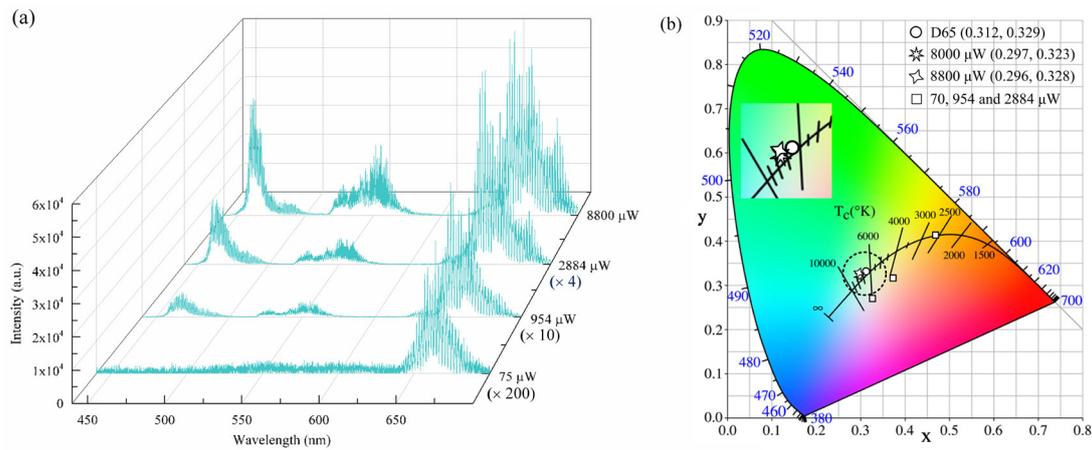

**Figure 3.** a) Lasing spectrum evolution with the increased pump power. b) Plot of the CIE coordinates of standard white illuminant D65 (0.312, 0.329) and lasing spectra under 8000-μW (0.312, 0.329) and 8800-μW (0.296, 0.328) pump powers.

Figure 3a presents the photoluminescence spectra of the microsphere cavity under different pump powers. To maintain the white laser under different pump powers in a single cavity, a dynamical balance between the optical gain and the feedback from the cavity for each color should be satisfied. Because the thresholds and the optical gains for three primary colors are different, the intensity proportion of three primary colors is variable as the pump power increases. It may be also used for the real-time laser color tuning. The spacings of the laser peaks for the red, green and blue lasers are 1.781, 1.28 and 0.86 nm, which are well matched with the theoretically calculated FSRs of 1.93,

1.241 and 0.896 nm. The microsphere cavity with a diameter of about 53 μm has abundant resonant modes in visible band, the competition between the lasing modes and the spontaneous emission is very intensive. Figures S2-4 show the detailed optical spectrum evolutions under the increased pump power, which is used to further confirm the lasing onset[33,42].

Under 8800-μW pump power, the RGB lasers are well balanced, and the calculated chromaticity (0.297, 0.323) is very closed to that of the CIE 1931 standard white illuminant D65 (0.312, 0.329)[43], as shown in Figure 3b. By further increasing the pump power to 8800 μW (0.296, 0.328), the CIE coordinate of the laser shifts little, as the RGB lasers are saturated. It is worth noting that, the blue laser is mainly contributed by the emission of Er ions, since the concentration of Tm ions is only a fiftieth of that of Er ions. However, the blue emission is ignorable in an Er-Yb codoped microsphere cavity, or a Tm-Yb codoped microsphere. It should be attributed to that the interaction between Er ions and Tm ions improves the blue emission from Er ions (The details are shown in the Figure S5).

Figure 4a shows the CCD image of the white-like lasing and its component RGB lasing in the microsphere cavity under 8000-μW pump power. However, it seems that the blue laser is dominated in this white-like lasing, while the green and red laser is relatively weak. This phenomenon should be concluded as the following three reasons: First, the response of our CCD at 670 nm is relatively weak. Second, there is a distinction between the emission intensities detected by the CCD and the spectrometer for these three lasing bands. Among the RGB lasers, the blue laser has the strongest optical scattering from the microsphere to free space for the CCD detection (generally, the shorter the wavelength is, the stronger scattering is), and the weakest coupling efficiency from the microsphere to the microfiber for the spectrometer detection (The theoretical result sees the Figure S6). Obviously, the blue emission is widely distributed in the whole microsphere due to the strongest scattering. Finally, due to the difference among the lasing wavelengths, the electrical field distributions of the RGB lasing modes are not fully overlapped. Figure 4b shows the electric field intensity distributions for the RGB lasing modes. As their maximum electric field intensity is located at the pumping position, $TE_{371}^{16}$, $TE_{439}^{19}$ and $TE_{535}^{22}$ modes should be attributed to the most excited modes for the red, green and blue lasers, respectively. Here, the superscript and subscript represent the axial and azimuthal quantum numbers, respectively. The $TE_{535}^{22}$ mode for the blue lasing has the largest axial and azimuthal quantum numbers, which means that the blue laser is homogeneously distributed with the smallest spatial spacing.

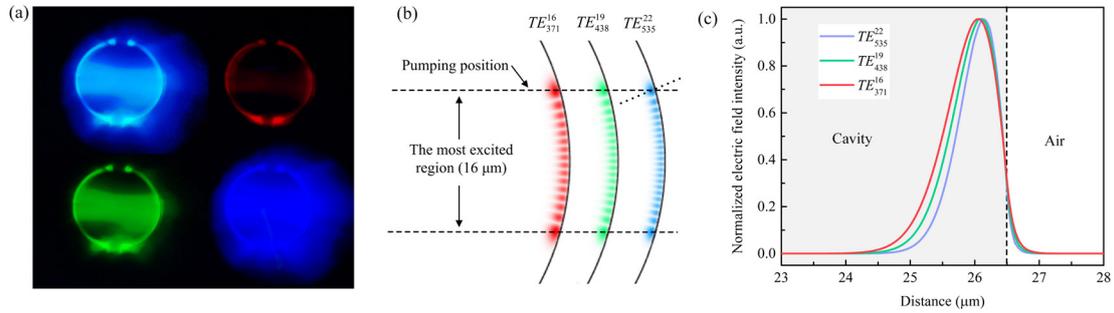

**Figure 4.** a) Optical microscope images of white-like lasing and its component RGB lasing. b) Electric field intensity distributions of $TE_{371}^{16}$, $TE_{439}^{19}$ and $TE_{535}^{22}$ modes on the cross-plane of the microsphere along the axis direction, which are calculated by the wavelengths of 460, 560 and 660 nm, respectively. The dotted line represents the radial direction, in which the detailed field profiles are shown in c).

The mode volumes of $TE^{16}_{371}$, $TE^{19}_{439}$ and $TE^{22}_{535}$ modes are estimated as 573, 487 and 399 μm³ respectively, indicating that the $TE^{22}_{535}$ mode with the smallest mode volume is well confined for the highest energy density. Figure 4c shows the mode field profiles along the radical direction. The peak intensities of $TE^{16}_{371}$, $TE^{19}_{439}$ and $TE^{22}_{535}$ modes are about 0.445, 0.412 and 0.372 μm away from the surface of the microsphere, respectively. It indicates that the $TE^{22}_{535}$ mode is more concentrated around cavity surface, which should lead to a stronger surface scattering light for the CCD detection. From this point of view, it is relatively random and difficult to extract the white laser by optical scattering. In contrast, it should be more flexible and controllable to drop the white laser by adjusting the coupling condition between the microsphere and the microfiber

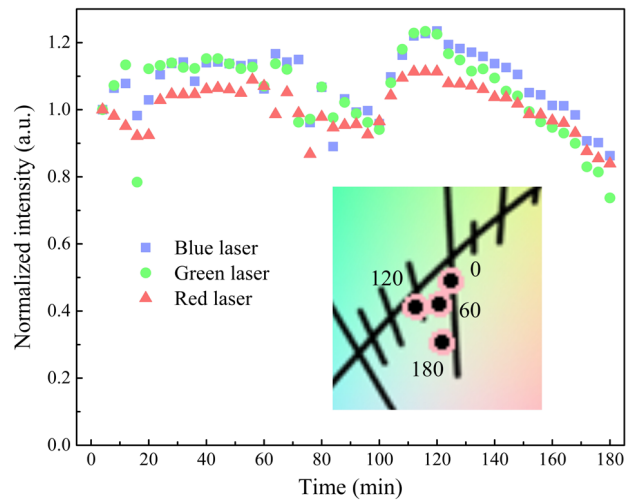

**Figure 5.** a) Lasing intensity evolutions of RGB over the duration of 180 minutes. The inset is the chromaticity of the laser at 0, 60, 120 and 180 minutes.

High-intensity continuous-wave pump usually causes thermal and optical damages for the gain materials, leading to the shift of the laser chromaticity and the deterioration of the emission intensity. We operated the Tm-Er-Yb codoped microcavity at room temperature for 180 min, under a 7-mW continuous-wave pump. As shown in Figure 5, the intensities of the RGB lasers vary within a range of 22%. This is a relatively large variation of the lasing intensity. However, their variations are synchronous. The inset of Figure 5 shows the shift of the chromaticity with a step of 60 minutes. It can be found that the shift is inconspicuous, and the chromaticity is always located at the white laser region, indicating a relatively good stability of the chromaticity.

**Conclusion**

In conclusion, we have demonstrated the CW upconversion RGB lasing in a Tm-Er-Yb co-doped microsphere cavity with an ultra-high-Q factor of $1.14\times10^8$, in which their lasing thresholds are as low as 90, 500 and 300 μW, respectively. As the pump power was increased to above 8000 μW, the emission intensities of the RGB lasers were well balanced, achieving the white laser in a single cavity. To our best knowledge, it is the first rare-earth based upconversion white microlaser under the room temperature and CW pump. Acquiring the RGB lasers from the microcavity by the optical scattering and the coupling waveguide are also discussed, in which the coupling waveguide is more controllable and extracts a higher proportion of the long-wavelength laser, compared with the optical scattering. The microsphere was continuously operated over 180 min, the chromaticity coordinate of the laser shifted barely, indicating a relatively stable white lasing. Our demonstration offers a simple, low-cost and eye-safe approach to realize the white microlaser by pumping mature near-infrared low-energy photons, which makes it suitable for applications in vivid display, holographic imaging and light-based version of Wi-Fi.

**Methods**

**Device fabrication.** The Tm-Er-Yb co-doped microsphere fabrication employs polymethyl methacrylate (PMMA) for doping rare earth elements without obvious ion clusters. First the Er-Yb-PMMA solution is made by mixing thulium nitrate hexahydrate (Tm $(NO_3)_3 \cdot 6H_2O$, 99.9%, Macklin), erbium nitrate pentahydrate (Er $(NO_3)_3 \cdot 5H_2O$, 99.9%, Macklin), ytterbium nitrate pentahydrate (Yb $(NO_3)_3 \cdot 5H_2O$, 99.99%, Macklin), polymethyl methacrylate (PMMA, TCI), and acetone (AR, Shanghai Hushi) with a weight ratio of 1:50:40:3105:81770. Second, the single-mode optical fiber was processed by $CO_2$ fabrication platform to form a microbottle cavity (the microsbottle fabrication detail see ref 34). Third, the microbottle was dipped into the prepared solution and then was drawn out. After evaporation, a Tm-Er-Yb-PMMA film will homogeneously form on the surface of the microbottle. Finally, the microbottle was reflowed to form a microsphere. During the reflow process, the PMMA will be fully eliminated and the rare earth element will be doped into the interior of the cavity.

**Device characterization.** The frequency detuning measurement is performed using a tunable diode laser (Toptica, CTL 1550) sweeping over the resonant modes. The CCD image is captured using an objective (0.28 numerical aperture ×20).

**Optical measurement.** A 975-nm CW semiconductor laser with the linewidth of about 1 nm was used to pump the Tm-Er-Yb codoped microsphere cavity via a tapered fiber. The emission light is coupled to a spectrometer (Horiba, iHR-550) by a collimator, and detected by an electric cooled silicon array charge-coupled device (Horiba, 1024X256-

OE).

**Data availability.** The data that supports the findings of this study is available from the corresponding author on reasonable request.

**Acknowledgments**

This work was supported by the National Natural Science Foundation of China (91850115, 11774110), the Fundamental Research Funds for the Central Universities (HUST: 2019kfyXKJC036, 2019kfyRCPY092), the State Key Laboratory of Advanced Optical Communication Systems and Networks (SJTU) (2021GZKF003), and the State Key Laboratory of Applied Optics (SKLAO2021001A10).

# Reference

1. Wang, T.; Yu, H.; Siu, C. K.; Qiu, J.; Xu, X.; Yu, S. F., White-Light Whispering-Gallery-Mode Lasing from Lanthanide-Doped Upconversion NaYF4 Hexagonal Microrods. *ACS Photonics* **2017,** *4* (6), 1539-1543.

2. Chen, Y.-C.; Tan, X.; Sun, Q.; Chen, Q.; Wang, W.; Fan, X., Laser-emission imaging of nuclear biomarkers for high-contrast cancer screening and immunodiagnosis. *Nature Biomedical Engineering* **2017,** *1* (9), 724-735.

3. Chen, Y.-C.; Fan, X., Biological Lasers for Biomedical Applications. *Advanced Optical Materials* **2019,** *7* (17), 1900377.

4. Lv, Y.; Li, Y. J.; Li, J.; Yan, Y.; Yao, J.; Zhao, Y. S., All-Color Subwavelength Output of Organic Flexible Microlasers. *J Am Chem Soc* **2017,** *139* (33), 11329-11332.

5. Zhao, J.; Yan, Y.; Gao, Z.; Du, Y.; Dong, H.; Yao, J.; Zhao, Y. S., Full-color laser displays based on organic printed microlaser arrays. *Nat Commun* **2019,** *10* (1), 870.

6. He, X.; Liu, P.; Wu, S.; Liao, Q.; Yao, J.; Fu, H., Multi-color perovskite nanowire lasers through kinetically controlled solution growth followed by gas-phase halide exchange. *Journal of Materials Chemistry C* **2017,** *5* (48), 12707-12713.

7. Xu, J.; Ma, L.; Guo, P.; Zhuang, X.; Zhu, X.; Hu, W.; Duan, X.; Pan, A., Room-temperature dual-wavelength lasing from single-nanoribbon lateral heterostructures. *J Am Chem Soc* **2012,** *134* (30), 12394-7.

8. Zhang, C.; Zou, C.-L.; Dong, H.; Yan, Y.; Yao, J.; Zhao, Y. S., Dual-color single-mode lasing in axially coupled organic nanowire resonators. *Science Advances* **2017,** *3* (7), e1700225.

9. Zhuang, X.; Ouyang, Y.; Wang, X.; Pan, A., Multicolor Semiconductor Lasers. *Advanced Optical Materials* **2019,** *7* (17).

10. Chang, S. W.; Liao, W. C.; Liao, Y. M.; Lin, H. I.; Lin, H. Y.; Lin, W. J.; Lin, S. Y.; Perumal, P.; Haider, G.; Tai, C. T.; Shen, K. C.; Chang, C. H.; Huang, Y. F.; Lin, T. Y.; Chen, Y. F., A White Random Laser. *Sci Rep* **2018,** *8* (1), 2720.

11. Liu, Z.; Yin, L.; Ning, H.; Yang, Z.; Tong, L.; Ning, C.-Z., Dynamical Color-Controllable Lasing with Extremely Wide Tuning Range from Red to Green in a Single Alloy Nanowire Using Nanoscale Manipulation. *Nano Letters* **2013,** *13* (10), 4945-4950.

12. Huang, D.; Xie, Y.; Lu, D.; Wang, Z.; Wang, J.; Yu, H.; Zhang, H., Demonstration of a White Laser with V2 C MXene-Based Quantum Dots. *Adv Mater* **2019,** *31* (24), e1901117.

13. Ding, Y.; Yang, Q.; Guo, X.; Wang, S.; Gu, F.; Fu, J.; Wan, Q.; Cheng, J.; Tong, L., Nanowires/microfiber hybrid structure multicolor laser. *Opt. Express* **2009,** *17* (24), 21813-21818.

14. Chen, S.; Zhao, X.; Wang, Y.; Shi, J.; Liu, D., White light emission with red-green-blue lasing action in a disordered system of nanoparticles. *Applied Physics Letters* **2012,** *101* (12).

15. Zhai, T.; Wang, Y.; Chen, L.; Wu, X.; Li, S.; Zhang, X., Red-green-blue laser emission from cascaded polymer membranes. *Nanoscale* **2015,** *7* (47), 19935-9.

16. Wang, C.; Gong, C.; Zhang, Y.; Qiao, Z.; Yuan, Z.; Gong, Y.; Chang, G.-E.; Tu, W.-C.; Chen, Y.-C., Programmable Rainbow-Colored Optofluidic Fiber Laser Encoded with Topologically Structured Chiral Droplets. *ACS Nano* **2021,** *15* (7), 11126-11136.

17. Shin, Y.; Park, S.; Kim, Y.; Lee, J., Development of a PC interface board for true color control using an Ar–Kr white-light laser. *Optics & Laser Technology* **2006,** *38* (4-6), 266-271.

18. Fan, F.; Turkdogan, S.; Liu, Z.; Shelhammer, D.; Ning, C. Z., A monolithic white laser. *Nat*

*Nanotechnol* **2015,** *10* (9), 796-803.

19. Zhou, B.; Shi, B.; Jin, D.; Liu, X., Controlling upconversion nanocrystals for emerging applications. *Nature Nanotechnology* **2015,** *10* (11), 924-936.

20. Chen, Z.; Dong, G.; Barillaro, G.; Qiu, J.; Yang, Z., Emerging and perspectives in microlasers based on rare-earth ions activated micro-/nanomaterials. *Progress in Materials Science* **2021**, 100814.

21. Deng, R.; Qin, F.; Chen, R.; Huang, W.; Hong, M.; Liu, X., Temporal full-colour tuning through non-steady-state upconversion. *Nature Nanotechnology* **2015,** *10* (3), 237-242.

22. Wang, F.; Han, Y.; Lim, C. S.; Lu, Y.; Wang, J.; Xu, J.; Chen, H.; Zhang, C.; Hong, M.; Liu, X., Simultaneous phase and size control of upconversion nanocrystals through lanthanide doping. *Nature* **2010,** *463* (7284), 1061-1065.

23. Höppe, H. A., Recent Developments in the Field of Inorganic Phosphors. *Angewandte Chemie International Edition* **2009,** *48* (20), 3572-3582.

24. Wang, F.; Deng, R.; Wang, J.; Wang, Q.; Han, Y.; Zhu, H.; Chen, X.; Liu, X., Tuning upconversion through energy migration in core–shell nanoparticles. *Nature Materials* **2011,** *10* (12), 968-973.

25. Liu, Y.; Lu, Y.; Yang, X.; Zheng, X.; Wen, S.; Wang, F.; Vidal, X.; Zhao, J.; Liu, D.; Zhou, Z.; Ma, C.; Zhou, J.; Piper, J. A.; Xi, P.; Jin, D., Amplified stimulated emission in upconversion nanoparticles for super-resolution nanoscopy. *Nature* **2017,** *543* (7644), 229-233.

26. Vicidomini, G.; Bianchini, P.; Diaspro, A., STED super-resolved microscopy. *Nature Methods* **2018,** *15* (3), 173-182.

27. Jiang, X.-F.; Zou, C.-L.; Wang, L.; Gong, Q.; Xiao, Y.-F., Whispering-gallery microcavities with unidirectional laser emission. *Laser & Photonics Reviews* **2016,** *10* (1), 40-61.

28. Godard, A., Infrared (2–12 μm) solid-state laser sources: a review. *Comptes Rendus Physique* **2007,** *8* (10), 1100-1128.

29. Haider, G.; Lin, H. I.; Yadav, K.; Shen, K. C.; Liao, Y. M.; Hu, H. W.; Roy, P. K.; Bera, K. P.; Lin, K. H.; Lee, H. M.; Chen, Y. T.; Chen, F. R.; Chen, Y. F., A Highly-Efficient Single Segment White Random Laser. *ACS Nano* **2018,** *12* (12), 11847-11859.

30. He, L.; Özdemir, Ş. K.; Yang, L., Whispering gallery microcavity lasers. *Laser & Photonics Reviews* **2013,** *7* (1), 60-82.

31. Yang, S.; Wang, Y.; Sun, H., Advances and Prospects for Whispering Gallery Mode Microcavities. *Advanced Optical Materials* **2015,** *3* (9), 1136-1162.

32. Vahala, K. J., Optical microcavities. *Nature* **2003,** *424* (6950), 839-846.

33. Kippenberg, T. J.; Kalkman, J.; Polman, A.; Vahala, K. J., Demonstration of an erbium-doped microdisk laser on a silicon chip. *Physical Review A* **2006,** *74* (5), 051802.

34. Zhu, S.; Shi, L.; Xiao, B.; Zhang, X.; Fan, X., All-Optical Tunable Microlaser Based on an Ultrahigh-Q Erbium-Doped Hybrid Microbottle Cavity. *ACS Photonics* **2018,** *5* (9), 3794-3800.

35. Lu, T.; Yang, L.; van Loon, R. V. A.; Polman, A.; Vahala, K. J., On-chip green silica upconversion microlaser. *Opt. Lett.* **2009,** *34* (4), 482-484.

36. Hsiu-Sheng, H.; Can, C.; Andrea, M. A. In *Low threshold Er3+/Yb3+ co-doped microcavity laser*, Proc.SPIE, 2010.

37. Sun, L.-D.; Dong, H.; Zhang, P.-Z.; Yan, C.-H., Upconversion of Rare Earth Nanomaterials. *Annual Review of Physical Chemistry* **2015,** *66* (1), 619-642.

38. Barnes, W. L.; Laming, R. I.; Tarbox, E. J.; Morkel, P. R., Absorption and emission cross section of Er/sup 3+/ doped silica fibers. *IEEE Journal of Quantum Electronics* **1991,** *27* (4), 1004-1010.

39. Gorodetsky, M. L.; Savchenkov, A. A.; Ilchenko, V. S., Ultimate Q of optical microsphere resonators. *Opt. Lett.* **1996,** *21* (7), 453-455.
40. Mehrabani, S.; Armani, A. M., Blue upconversion laser based on thulium-doped silica microcavity. *Opt. Lett.* **2013,** *38* (21), 4346-4349.
41. Fernandez-Bravo, A.; Yao, K.; Barnard, E. S.; Borys, N. J.; Levy, E. S.; Tian, B.; Tajon, C. A.; Moretti, L.; Altoe, M. V.; Aloni, S.; Beketayev, K.; Scotognella, F.; Cohen, B. E.; Chan, E. M.; Schuck, P. J., Continuous-wave upconverting nanoparticle microlasers. *Nature Nanotechnology* **2018,** *13* (7), 572-577.
42. Liang, Y.; Zhu, H.; Zheng, H.; Tang, Z.; Wang, Y.; Wei, H.; Hong, R.; Gui, X.; Shen, Y., Competition of whispering gallery lasing modes in microwire with hexagonal cavity. *Journal of Physics D: Applied Physics* **2020,** *54* (5), 055107.
43. Luo, Z.; Chen, Y.; Wu, S. T., Wide color gamut LCD with a quantum dot backlight. *Opt Express* **2013,** *21* (22), 26269-84.